\documentclass[preprint]{aastex}
\usepackage{graphicx}

\begin{document}


\title{\bf Effects of variable eccentricity on the climate of an Earth-like world}

\author{M.J. Way\altaffilmark{1}}
\affil{NASA Goddard Institute for Space Studies, 2880 Broadway, New York, NY 10025, USA}
\email{Michael.J.Way@nasa.gov}
\and

\author{Nikolaos Georgakarakos}
\affil{New York University Abu Dhabi, Saadiyat Island, PO Box 129188, Abu Dhabi, UAE}

\altaffiltext{1}{Department of Physics and Astronomy, Uppsala University, Uppsala, 75120, Sweden}


\begin{abstract}
The Kepler era of exoplanetary discovery has presented the Astronomical
community with a cornucopia of planetary systems very different from the one
which we inhabit.  It has long been known that Jupiter plays a major role in
the orbital parameters of Mars and it's climate, but there is also a
long-standing belief that Jupiter would play a similar role for Earth if not
for its large moon. Using a three dimensional general circulation model (3-D
GCM) with a fully-coupled ocean we simulate what would happen to the climate of
an Earth-like world if Mars did not exist, but a Jupiter-like planet was much
closer to Earth's orbit.  We investigate two scenarios that involve evolution
of the Earth-like planet's orbital eccentricity from 0--0.283 over 6500 years,
and from 0--0.066 on a time scale of 4500 years.
In both cases we discover that they would maintain relatively temperate climates
over the time-scales simulated. More Earth-like planets in multi-planet
systems will be discovered as we continue to survey the skies and the results
herein show that the proximity of large gas giant planets may play an important
role in the habitability of these worlds.  These are the first such 3-D GCM
simulations using a fully-coupled ocean with a planetary orbit that evolves
over time due to the presence of a giant planet.

\end{abstract}

\section{Introduction} \label{Introduction}

The Kepler era has demonstrated a plethora of planetary systems whose orbital
configurations are quite unlike those of our own solar
system.\footnote{http://exoplanet.eu/diagrams/} The discovery of
planets of similar mass and size to those of the Earth has ignited a 
strong interest in modeling the possibility of worlds
in such system's habitable zones (HZ). As of November 2016 there are 31 planets\footnote{http://exoplanet.eu/}
with masses less than 2.5 M$_{\earth}$ with eccentricities larger than 0.1, albeit
with relatively large errors bars in each quantity for most objects. Of these 6 are
in multi-planet systems. Using a different criteria \cite{Bolmont2016} find
four possible terrestrial type planets with a non-negligible percentage of
their orbit in the HZ for eccentricities greater than 0.1.
More such systems will be discovered in the future and it is likely that there
will exist cases where a Jupiter-like planet may be influencing the
orbital parameters of an Earth-like terrestrial world more so than in the
current day Earth-Jupiter case. Understanding variations such as
eccentricity and polar obliquity ($\theta_{p}$) will be important for the
climate and habitability potential of a terrestrial planet in
such a system.  For example, the effect Jupiter has on the eccentricity and
obliquity states of Mars are well known
\citep{Ward1991,Laskar2004,Armstrong2004}, but Jupiter also has an effect on
Earth \citep[][hereafter SP2010]{Spiegel2010} which we see as part of
the Milankovich cycles \citep{Milankovich1941}. At the same time some studies
show that Earth's large moon offers it more obliquity stability against
Jupiter's influence \citep[e.g.][]{Laskar1993} and may be a requirement for climate
stability and life \citep[e.g.][]{Waltham2004}, although \cite{Lissauer2011}
claim it is not as important as previously thought.  In fact Jupiter plays a
role in the present-day orbital dynamics of Venus, Earth and Mars,
but it may also have played a role in their
formation and size \citep[e.g.][]{Fritz2014,Batygin2015}. This will be no less true
for similar or more ``tightly packed" extrasolar planetary systems.

The effects of the orbital eccentricity and the obliquity of an Earth-like
planet have been investigated in several studies. \cite{WilliamsPollard2002} (hereafter WP2002)
used a 3-D General Circulation Model (GCM) and a 1-D Energy Balance Model (EBM)
to show that the ``average stellar flux received over an entire orbit, not the
length of the time spent within the HZ'' determine the long-term climate
stability of systems with a broad range of eccentricities (0.1--0.7). All of their
simulations used a semi-major axis of 1 Astronomical Unit (AU) for the
Earth-like planet. They found that one avoids reaching the moist greenhouse
and runaway greenhouse limits as long as eccentricities are less than 0.42 and 0.70
respectively. A 3-D GCM was also used by the same authors for investigating the effects
that various $\theta_{p}$ angles may have on the climate of an Earth-like planet
\citep{WilliamsPollard2003}.  They concluded that most Earth-like planets should
be hospitable to life at high obliquity.  None of their simulated planets were
warm enough to develop a runaway greenhouse or cold enough to freeze over
completely. The recent study by \cite{Bolmont2016}, while not directly comparable
to that herein because of their model choices, is nonetheless interesting.
They investigated the effectiveness of the mean flux approximation
as previously studied by WP2002. However, their setup was distinct:
they ran 3-D GCM simulations using 3 different stellar luminosities
(L$_{\star}$=L$_{\sun}$,10$^{-2}$L$_{\sun}$, 10$^{-4}$L$_{\sun}$),
with a range of semi-major axes, fixed eccentricities and
orbital periods while keeping $\theta_{p}$=0.
All systems were in a 1:1 spin-orbit resonance. They concluded 
``that the higher the eccentricity and the higher the luminosity of the star,
the less reliable the mean flux approximation."

\citet{Dressing2010} (hereafter D2010) used an EBM to examine
how different fixed values of eccentricity, $\theta_{p}$, azimuthal
obliquity, ocean fraction, and rotation rate of a terrestrial type world might
affect climate. They also explored the transition to a snowball state by
reducing the stellar luminosity.  They found that the polar regions at higher
eccentricities (where $\theta_{p}$=23.5$\arcdeg$ was kept fixed) receive more
mean insolation and the regional habitability fraction increased.
The result is clear: if Earth had a larger orbital
eccentricity it may have had even larger areas of habitability than it does
today (see Section \ref{Discussion} for direct comparison to this work).  They
also showed that the outer edge of the HZ expands with values of
eccentricity 0.4-0.7.  The findings of D2010 confirm the GCM results of WP2002
in that increasing the eccentricity of a terrestrial world can increase the
allowed semi-major axis for the outer edge of the HZ.  D2010 also
note that as one increases eccentricity regional and seasonal variability also
increase in amplitude leading to ``a more gradual transition from habitable to
non-habitable planets with increasing semi-major axis.''

A companion paper to that of D2010 is by SP2010 who use the same EBM as in
D2010, but have no 3-D GCM simulations.  The focus in SP2010 is slightly
different as much of the paper examines what sorts of eccentricities are
required to `break out' of a cold-start condition like that of a snowball
state. At the same time it is one of the few papers to consider the effects of
{\em variable eccentricity} on long-term climatic habitability.

Other recent studies regarding habitable worlds include \cite{Linsenmeier2015}
who used a GCM to explore the effects of seasonal variability for the climate
of Earth-like planets as determined by $\theta_{p}$ and orbital eccentricity
and \cite{Armstrong2014} who used an EBM to study the impact of obliquity
variations on planetary habitability in hypothetical systems. 
\cite{Ferreira2014} use a 3-D GCM to investigate high obliquity states with a fully
coupled ocean in the context of an aquaworld. They explore three obliquities (23.5, 54 and 90{\arcdeg})
and find in all cases that their world still appears to be habitable.

In this work, we investigate the effects on the climate of an Earth-like planet
whose orbit is perturbed by the presence of a nearby giant planet. For the first
time, a GCM coupled with analytical equations that describe the orbital
evolution of a terrestrial planet are used.  An additional major difference
between our work and previous studies is that 
we utilize a fully-coupled ocean model and an Earth continental layout. This is in
contrast to WP2002 who used a 50 meter ``thermodynamic slab" ocean model without
horizontal ocean heat transport or \cite{Linsenmeier2015} who used an aquaplanet model
and a 50m slab ocean, but again with no horizontal ocean heat transport.
We use a fully-coupled ocean model because
alongside atmospheric heat transport, ocean heat transport plays a vital role
in the climate of Earth \citep{Peixoto1992}.  In particular the work of
\cite{HuYang2014} has shown that the effects of a fully coupled ocean versus a
shallow slab ocean can be significant when looking at syncrononously rotating
worlds around M-dwarf stars. \cite{Godolt2015} demonstrated stark differences
for planets orbiting F-type stars when changing ocean heat transport while
\cite{Rose2015} has nicely demonstrated the climatic effects of changing ocean heat transport
equations for aqua and ridge type worlds.  The downside of a fully-coupled ocean approach is
that it can take hundreds of model years for a fully-coupled ocean to come into
equilibrium with the atmosphere, yet it will provide a more accurate picture of
the climate of the world being modeled.  We focus this study on the effects that the
terrestrial planet's orbital eccentricity has on the planet's climate, which is
an under-researched area in 3-D GCM studies. At the same time we keep
$\theta_{p}$=23.5$\arcdeg$ as for modern Earth.  The latter is a necessary
requirement for comparing with past and future work in the literature since
obliquity plays such an important role in the possible climate states of
terrestrial planets.

\section{Methods} \label{Methods}

Our terrestrial planet climate simulations utilize the Goddard Institute for Space
Studies 3-D GCM known as Model E2 \citep{Schmidt2014}. The version used
for this work is referred to as ROCKE3D\footnote{Resolving Orbital and Climate
Keys of Earth and Extraterrestrial Environments with Dynamics} \citep{Way2017}.
ROCKE3D has extensions to Model E2 to allow for a larger range of temperatures,
different atmospheric constituents, topographies, rotation rates and variable
eccentricity in time.  An Earth-like topography is used for the atmospheric
simulations herein, but with modest changes from present day Earth to make
the model more robust to possibly extreme conditions encountered by the perturbed orbits
modeled.  The model is run on a 4x5$\arcdeg$ latitude-longitude grid with 20
vertical atmospheric layers with the top set to 0.1hPa. A 13 layer fully
coupled ocean is utilized, but is a simplified `bathtub' type of ocean
topography with depths along coasts of 591 meters, and 1360 meters
elsewhere.\footnote{The ocean depth numbers correspond to specific ocean layers
in the model.} The shallower ocean allows the model to attain thermodynamic
equilibrium more quickly than it would if using an actual Earth $\sim$5000 meter
depth ocean. Some shallow basins were also filled in such as Hudson Bay, the
Mediterranean, the Baltic Sea, and the Black Sea. A number of straits were opened
up, like that north of Baffin Bay and those north of Australia.
The same solar insolation at 1AU used for Earth
studies is used here.  The atmospheric constituents and green house gas amounts
are the same as modern day Earth with 984mb of atmospheric pressure at the surface.
The $\theta_{p}$ angle and rotation rate are also the same as modern day Earth.
ROCKE3D does not take into account the resolved effects of non-dilute species,
although it does take this into account in determining parcel buoyancy in the
cumulus parameterization. Water vapor is the largest potential problem in this regard.
However, not until the water vapor becomes greater than 10-15\% of the mass
does it start to become problematic as pointed out in \cite{PD2016} and we do not reach
this limit in any of the simulations presented herein (see Section \ref{Discussion}).

In order to model the orbital evolution of the Earth-like planet, we make
use of the work by \cite{Georgakarakos2016}. This
study, which builds on previous results \citep[e.g.][]{Georgakarakos2003,
Georgakarakos2015}, focuses on the long term orbital evolution of a terrestrial
planet under the gravitational influence of a Jupiter-like world with
the latter relatively close to the former.  The analytical equations for the
orbital motion of the Earth-like planet derived in \cite{Georgakarakos2016}
are used to provide eccentricity and pericentre values for the 3-D
GCM simulations herein. In the context of this work, all bodies
are treated as point masses, they lie on the same plane of motion and the
system is not close to a mean motion resonance.

We simulated two different climate evolution scenarios as outlined in Table
\ref{table:table1}.  Unlike the previous studies mentioned in the Introduction
we were not able to run as many parameters ensembles given the length of time
it takes the GCM to run one secular period of the Earth-like planet's motion
(see column 9 in Table \ref{table:table1}).  The GCM initial conditions are the same as
modern Earth surface and ocean temperature values, unlike that of SP2010 who start many of
their simulations in a snowball state and then examine how differing
eccentricities might pull the world out of that state and into a possibly more
temperate one. As mentioned above, we maintain a fixed Earth-like polar obliquity
($\theta_{p}$=$\theta_{\earth}$) to make it more directly comparable to modern
day Earth.  Certainly a broader range of obliquities need to be considered in
the future, not to mention the possible coupling of variable obliquity with variable
eccentricity whose values are driven by work similar to that in \cite{Georgakarakos2016}.

\noindent
\begin{table}
\centering
\caption{Simulations}\label{table:table1}
\resizebox{\textwidth}{!}{
\begin{tabular}{|c|c|c|c|c|c|c|c|c|} \hline
Case  & \multicolumn{2}{c|}{Jupiter} & \multicolumn{4}{c|}{Earth}  & \multicolumn{2}{c|}{runtime} \\
\hline
      & eccentricity & semimajor axis & eccentricity & obliquity  & orbital/rotation period & semimajor axis & model years & wall clock \\
\hline
 1    &  0.27        &2.15            & variable     & 23$\arcdeg$& 365d/24hr               & 1.00   &         7000  &       108 days\\
 2    &  0.05        &1.80            & variable     & 23$\arcdeg$& 365d/24hr               & 1.00   &         5000  &       62 days\\ 
\hline
\end{tabular}}
\end{table}


\section{Results and Discussion}\label{Discussion}
\begin{figure}[ht!]
\figurenum{1}
\centering
\includegraphics[scale=0.4]{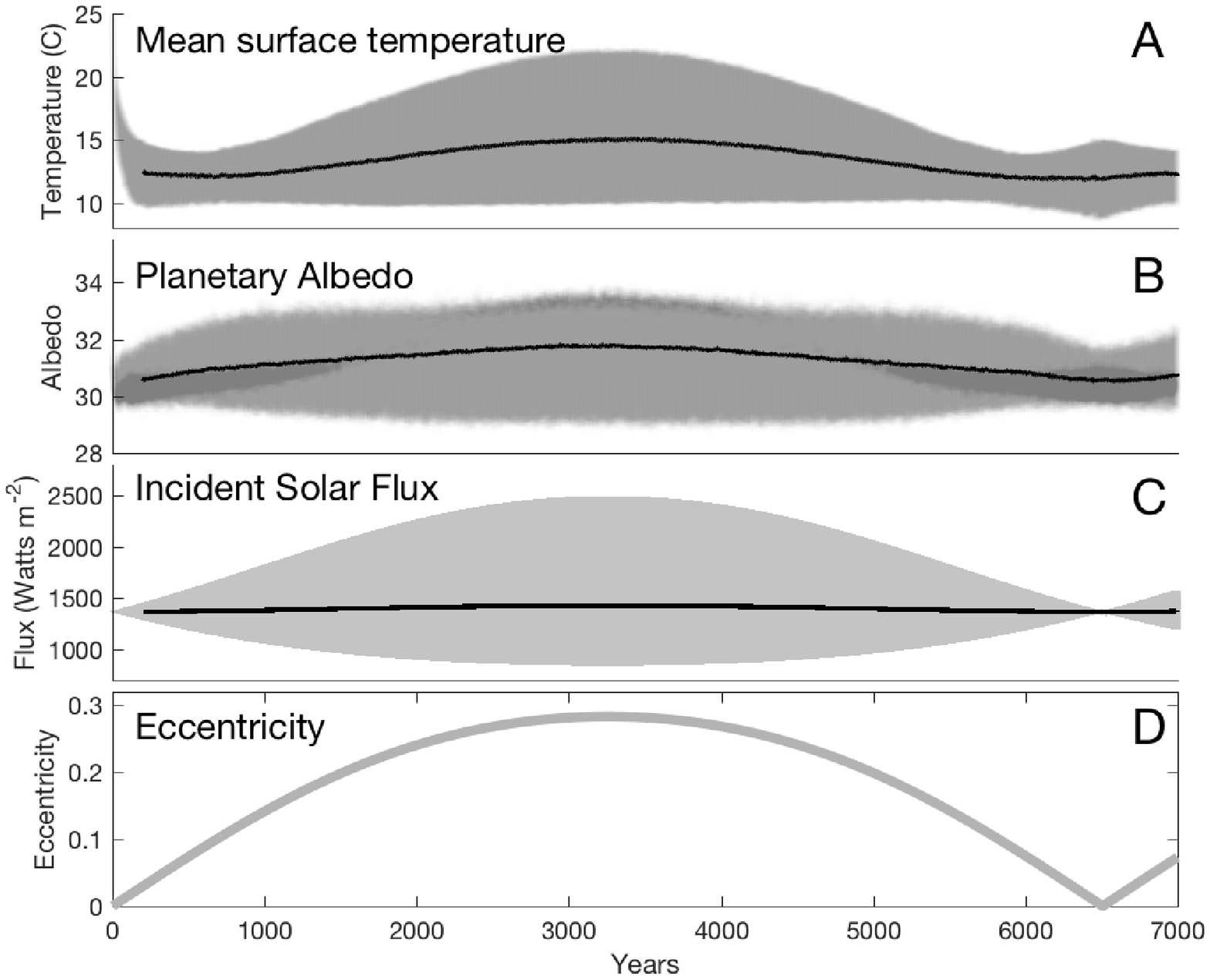}
\includegraphics[scale=0.4]{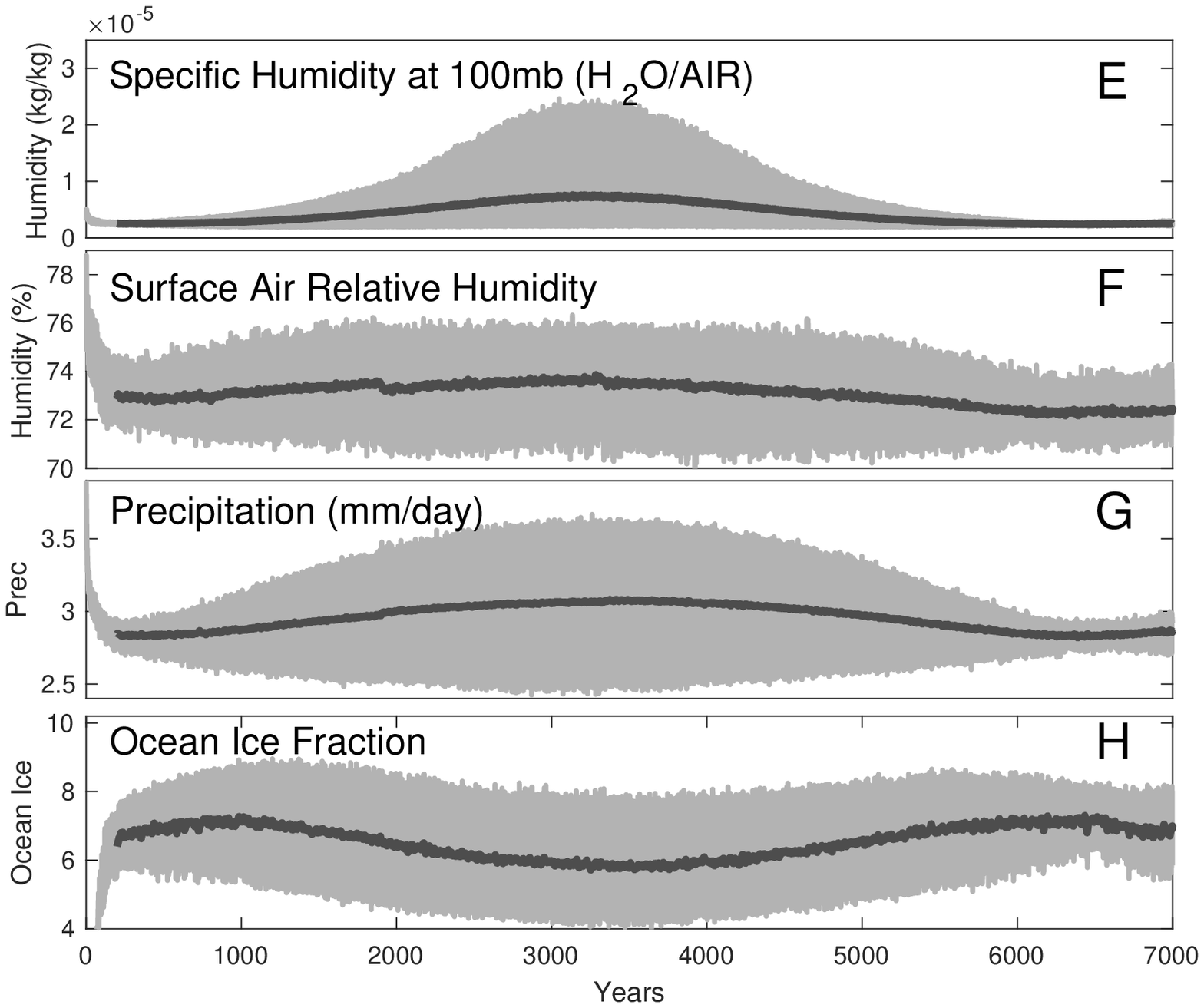}
\caption{Case 1 from Table 1. Solid black lines are a 10 year running mean. 
Plot A is the mean {\em Surface Air Temperature} for the Earth-like planet. B
is the {\em Planetary Albedo} in percentage. C is the amount of {\em Incident
Solar Flux} the planet receives in Watts per meter squared.  D is the {\em
Orbital Eccentricity} as a function of time.  Plot E is the globally averaged
{\em Specific Humidity} at 100mb (the top layer of our GCM atmosphere) given in
units of kilograms of H$_{2}$O per kilogram of air.  F is the surface air {\em
Relative Humidity} (the lowest atmospheric layer in our GCM).  G is the {\em
Precipitation} in mm per day, while plot F is the {\em Ocean Ice Fraction}
defined as the percentage of the ocean that is covered in ice.}\label{figure:1}
\end{figure}

\begin{figure}[ht!]
\figurenum{2}
\centering
\includegraphics[scale=0.4]{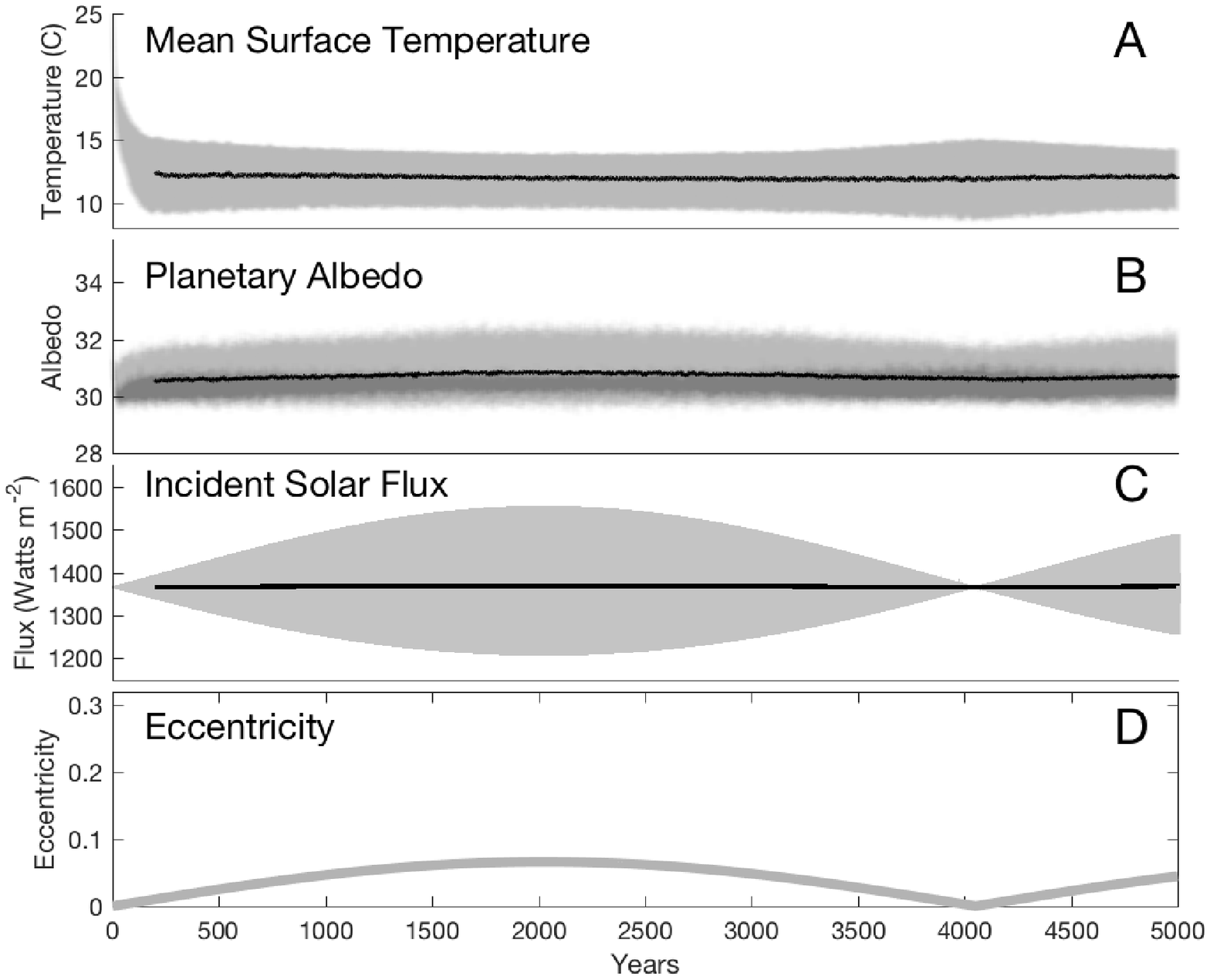}
\includegraphics[scale=0.4]{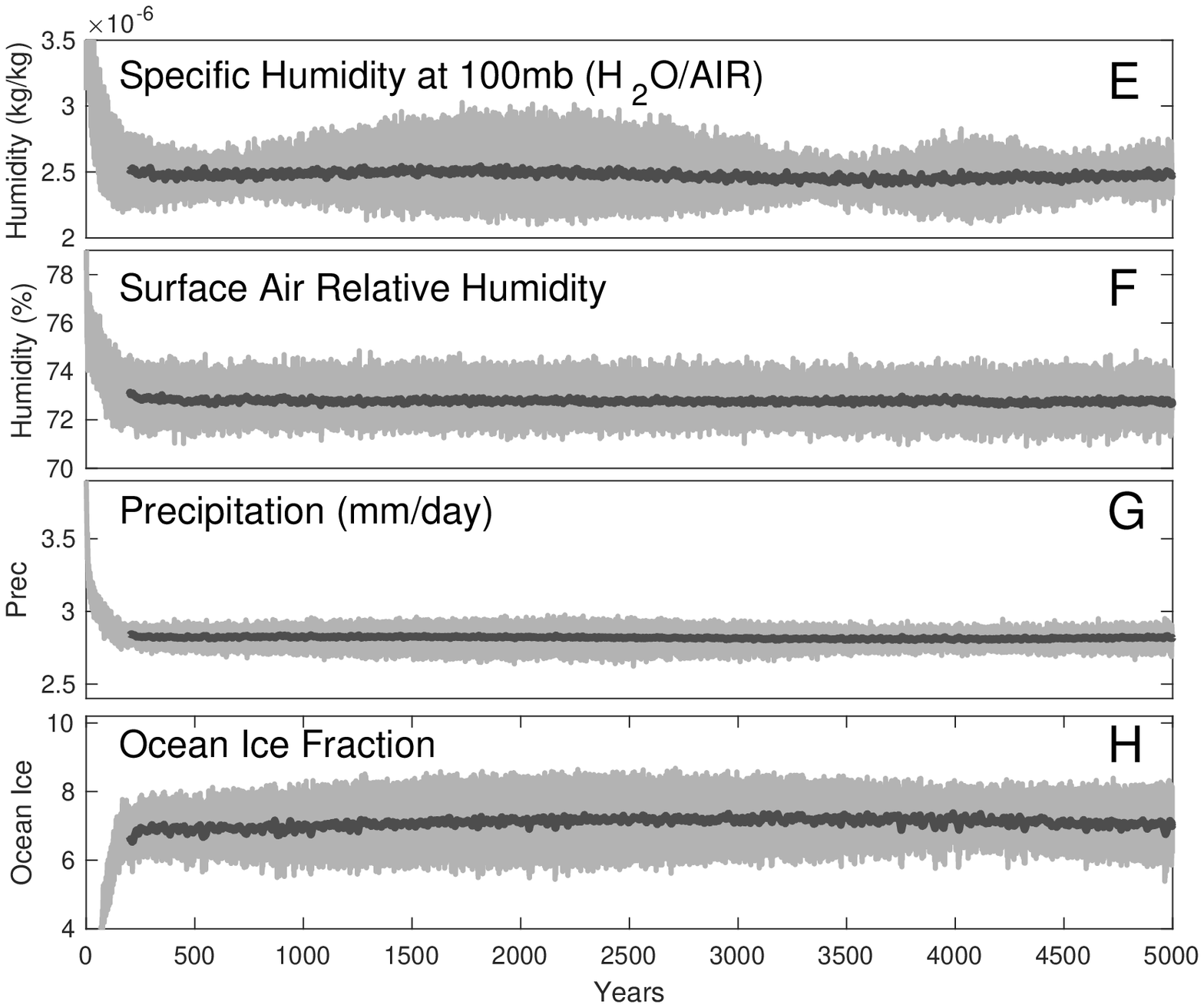}
\caption{Case 2 from Table 1. All ordinates are the same as for Case 1 except
for plots C and E that have been adusted to allow for more visible
detail.}\label{figure:2}
\end{figure}

There are a number of interesting similarities and differences between the results
herein and that of WP2002 (the work that most closely corresponds to ours).  We
show, as demonstrated in WP2002, that relative humidity and precipitation
increase with increasing temperature (Figures \ref{figure:1} and \ref{figure:2}
plots F and G) when the planet is near periastron for a given orbit at the
higher eccentricities shown in Figures \ref{figure:1}D and \ref{figure:2}D.

\begin{figure}[ht!]
\figurenum{3}
\centering
\includegraphics[scale=0.27]{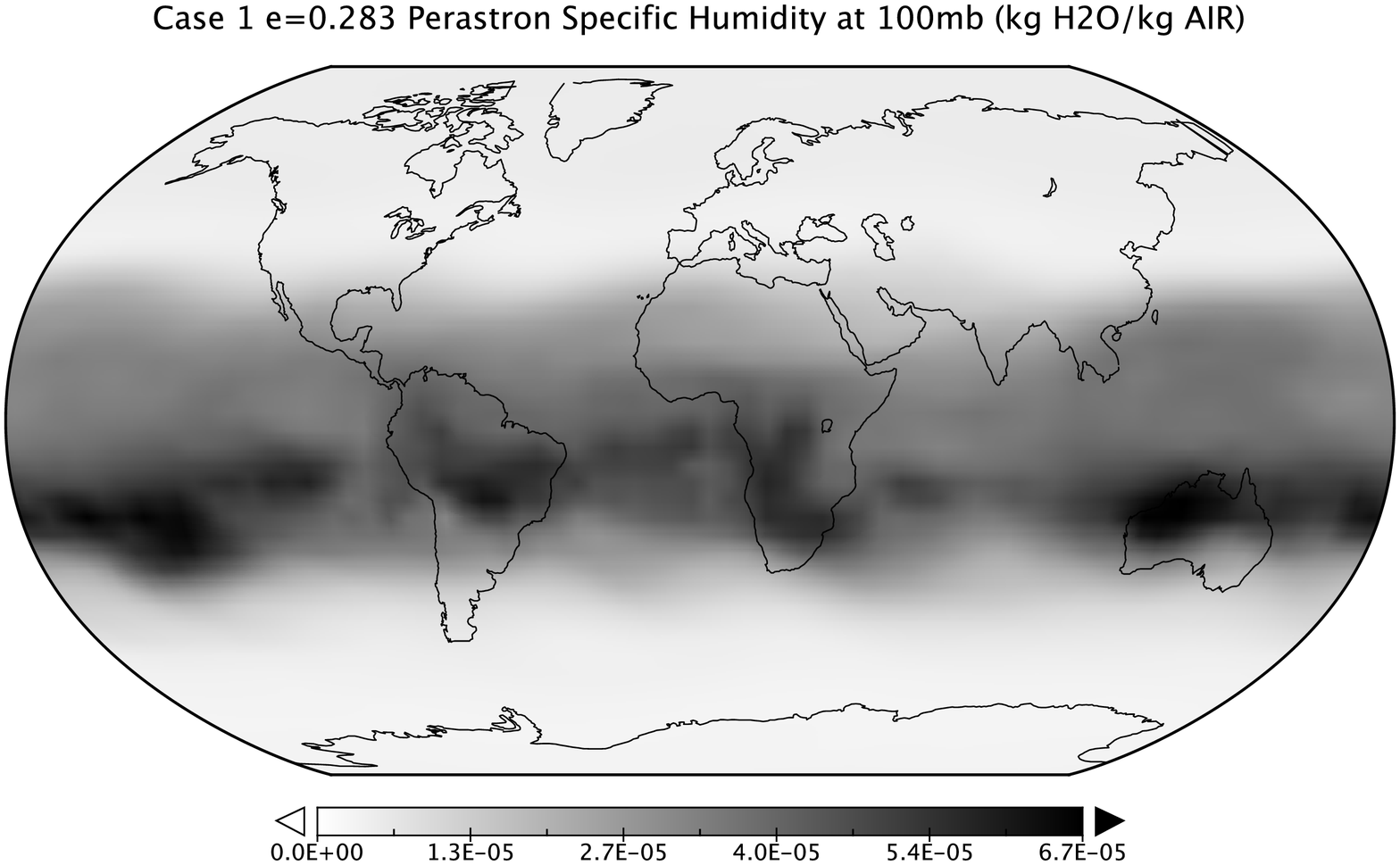}
\includegraphics[scale=0.27]{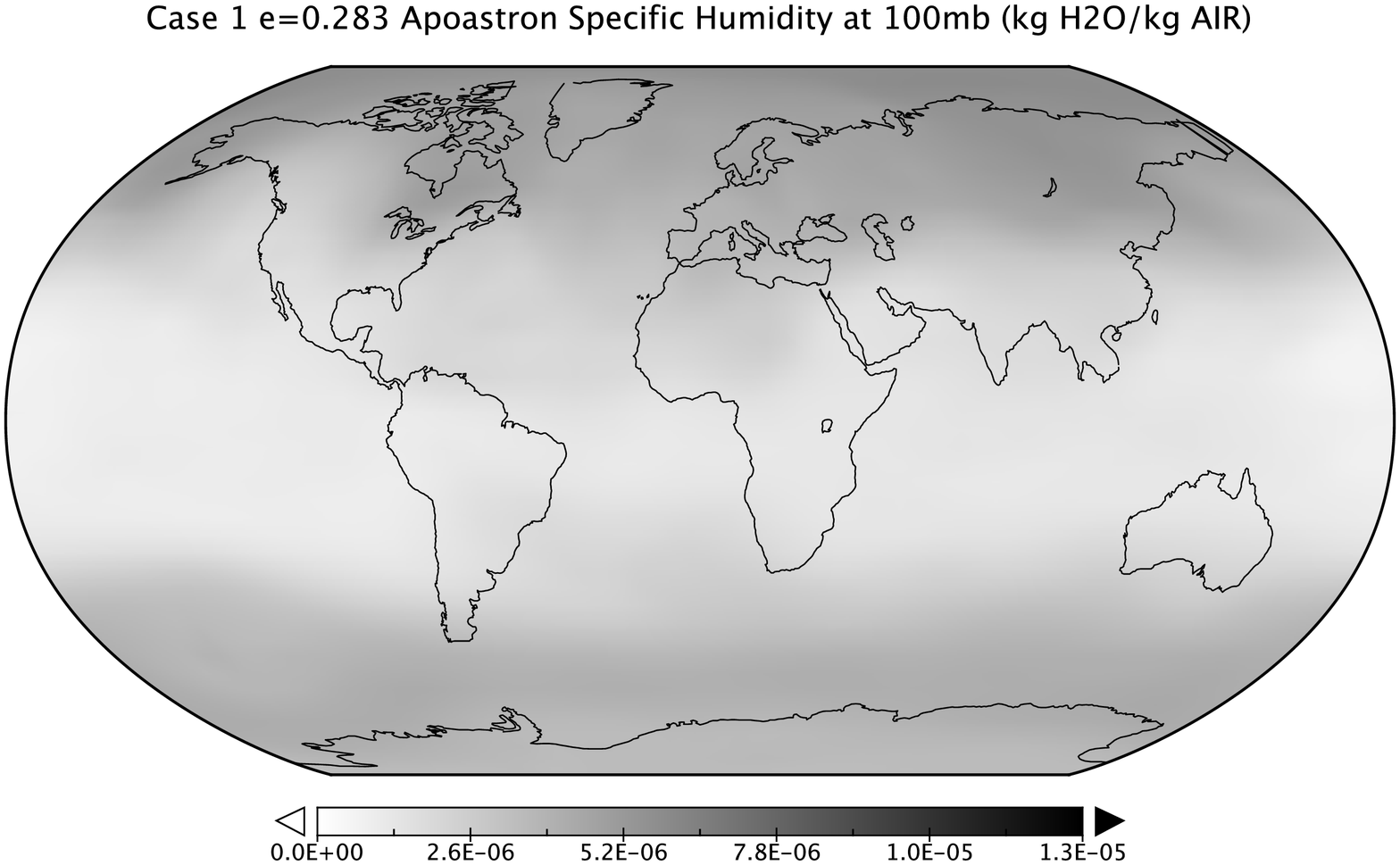}
\caption{Case 1 Specific Humidity at 100mb at Periastron (left) and Apoastron (right).
Note that the scale of the figures is different. The Apoastron limits are 1/5 the limits
of those for Periastron.}\label{figure:3}
\end{figure}

\begin{figure}[ht!]
\figurenum{4}
\centering
\includegraphics[scale=0.27]{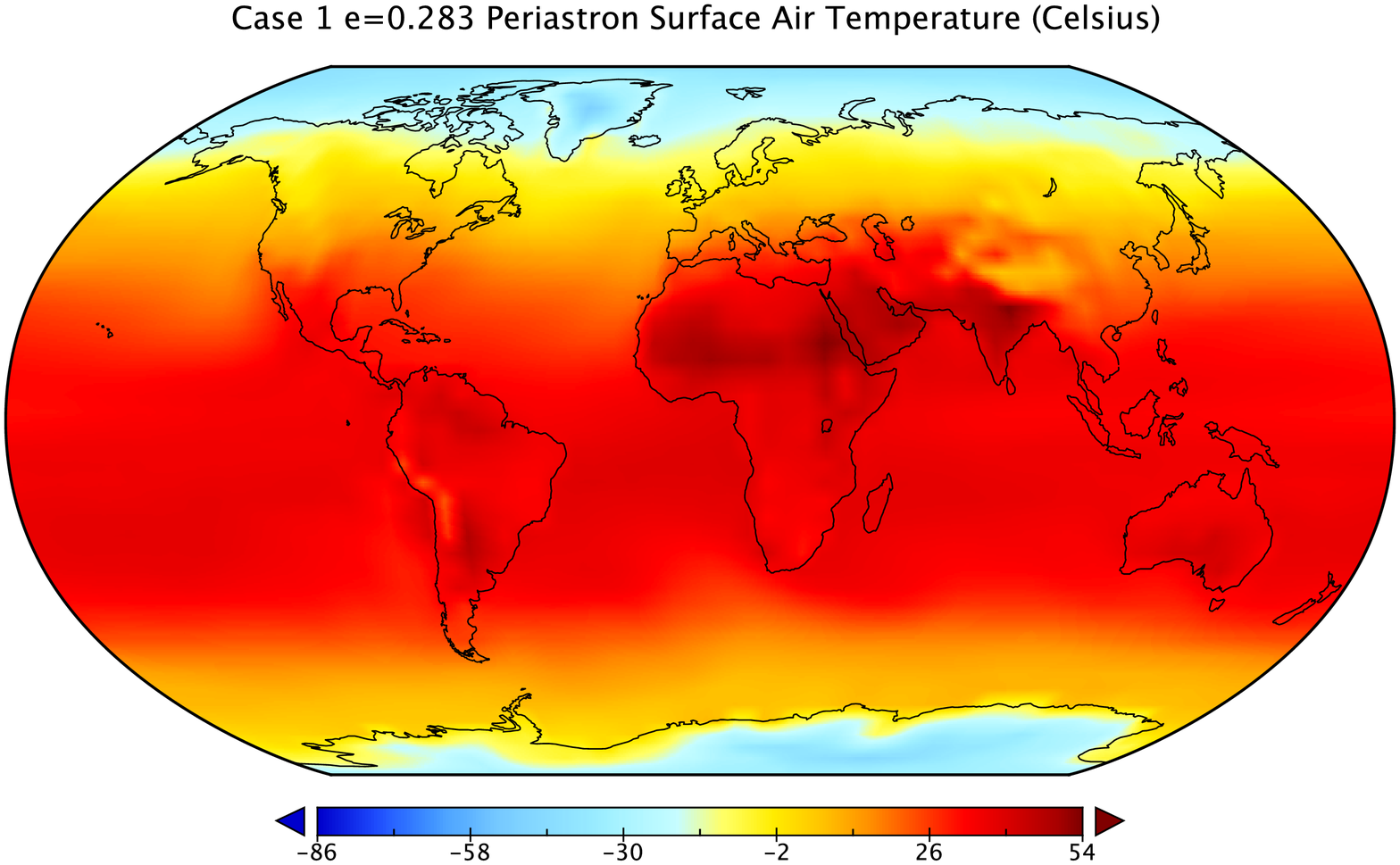}
\includegraphics[scale=0.27]{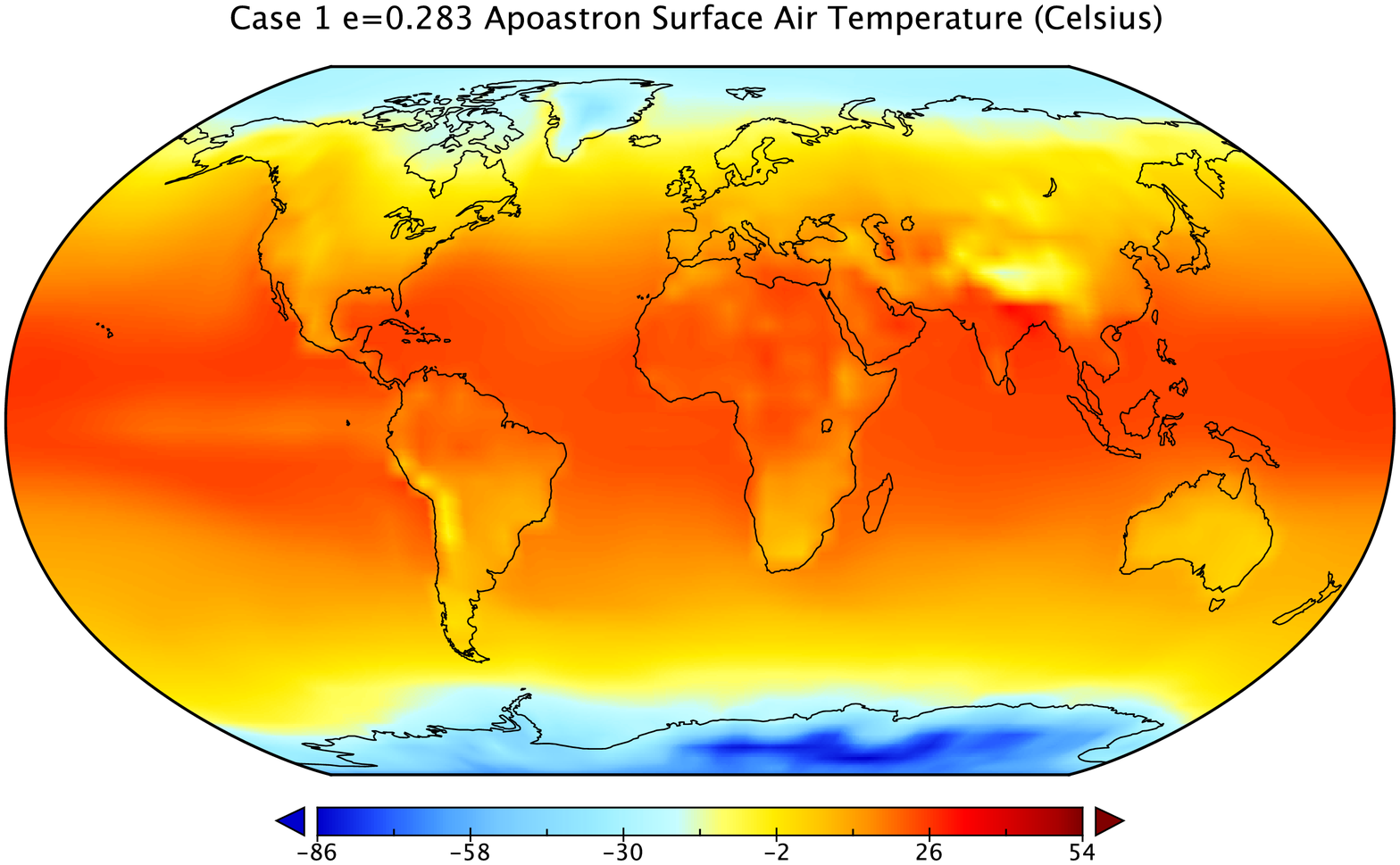}
\caption{Case 1 surface temperatures at Periastron (left) and Apoastron (right).
The gridding corresponds to the GCM lat-lon resolution used (4x5$\arcdeg$).}\label{figure:4}
\end{figure}

These effects are more pronounced for Case 1 (shown in Figure \ref{figure:1})
because of the higher eccentricities achieved. The planetary mean relative
humidity for the first layer of the atmosphere for Case 1 as shown in Figure
\ref{figure:1}F when the planet is near periastron for a given orbit is quite a
bit higher than present day Earth ($\sim$73\%). The precipitation in Figure \ref{figure:1}G
goes to higher values than the average modern Earth value of $\sim$3 mm/day \citep{Legates2008}.
Higher than modern day Earth specific humidities in the top atmospheric layer
(less than 2x10$^{-6}$kg H$_{2}$O / kg AIR) for Case 1 also manifest themselves
in Figure \ref{figure:1}E and Figure \ref{figure:3}A, but they are two orders of magnitude below the
\cite{Kasting1988} moist-greenhouse limit (the ``Kasting limit") near
each periastron crossing when the eccentricities are at their highest.  It
should be noted that at model year 3359, when the highest eccentricity and
solar insolation is reached at periastron, the max gridpoint temperature was
53.9$\arcdeg$C (see Figure \ref{figure:4}A). This is within the range of validity of the ROCKE3D GCM
radiation scheme.  WP2002 state that their eccentricity limit for a
moist-greenhouse state is 0.42, but in Case 1 (Figure \ref{figure:1}E and \ref{figure:3}A) it is clear that at
our max eccentricity of 0.283 we are very var from approaching a
moist-greenhouse state.  In neither Case 1 or 2 are these worlds anywhere near
the moist greenouse limit as evidenced in Figure \ref{figure:1}E and \ref{figure:2}E.
No individual grid cell in either case approaches the moist greenhouse limit at
periastron. For Case 1 the highest monthly
averaged specific humidity achieved in a given grid cell at 100mb at
periastron for the highest eccentricity achieved is
6.7x10$^{-5}$ kg H$_{2}$O / kg AIR (Figure \ref{figure:3}A). While this is nearly 3 times the max globally
averaged monthly value plotted in Figure 1E, it is still nearly 1.5 orders of
magnitude lower than the moist greenhouse limit. As mentioned in Section
\ref{Methods} ROCKE3D does not into account the resolved effects of non-dilute
species.  For Case 1 we quantified this effect for water vapor at the highest
eccentricity achieved, 0.283.  We looked at 30 min intervals (our Physics
time-step) over the month around periastron. The highest surface specific
humidity found was nearly 0.048 kg H$_{2}$O/kg AIR. This corresponds to less
than 5\% of the mass of the atmosphere, and therefore should not be a significant
source of error.

The closest simulation in WP2002 to our Case 1 at maximum eccentricity is their
run ``GCM 2" which has a fixed eccentricity of 0.3.  Their mean surface
temperature is 22.90$\arcdeg$C, which is very close to our Case 1 22.5$\arcdeg$C
for our largest eccentricity near orbital periastron.
This gives us confidence that our results are consistent with WP2002 in some ways.

As in WP2002 our albedo and ocean ice fraction in Case 1 (Figure
\ref{figure:1}B and \ref{figure:1}H) show some interesting behavior (it is more
difficult to discern in Case 2). At higher eccentricities the ocean ice
fraction decreases at the same time that the spread in albedos is largest.
Surely this is related to the world's ability to keep the ocean ice fraction
low at high eccentricities even at apoastron. This is likely a side effect of
the bulk heat capacity of our temperate world \citep{Cowan2012}. It is also due
to our fully coupled ocean's horizontal heat transport and the relatively short
amount of time Case 1 spends at its farthest extent from the Sun.

D2010 also discuss the importance of model relaxation in their Appendix.  We
have also tested this, which is necessary given that we have a fully coupled
ocean rather than their shallow 50 meter slab ocean. Using a 10 year running
mean we found that the net radiative balance for our world was within +/- 0.2
Watts m$^{-2}$ for the entirety of Simulation 1 and 2. This demonstrates that
even for our most rapidly changing world (Simulation 1) it is not changing fast
enough to throw the planet's radiative balance off enough to affect our
results.

\section{Conclusion}\label{Conclusion}

With upcoming space missions such as the Transiting Exoplanet Survey Satellite
\citep{Ricker2014}, the James Webb Space Telescope \citep{Gardner2006} and large
ground based observatories such as the European Extremely Large Telescope 
\citep{Gilmozzi2007} systems similar to those described herein are likely to be discovered.
Prior to follow-up observations, which may be costly in terms of telescope observing time,
it will be important to contrain any evolution in the orbital parameters of 
worlds like those of Case 1 whose regional habitability may be larger than other potential
candidates. We have not discussed the evolution of polar orbital
inclination of the Earth-like planet in this work, but other 3-D GCM studies (e.g.
\cite{WilliamsPollard2003,Linsenmeier2015}) have shown that it plays an
additional role in the habitability states of worlds like those we have modeled
in this work. Future studies in this series will include polar orbital inclination evolution
to better understand its effects in combination with eccentricity. Rotation rate will
also be examined since it is clear that it plays a very important role in
understanding the extent of the HZ \citep{Iro2010,Yang2014}.

\acknowledgements
Thanks goes to Thomas P. Clune for his help in getting the ROCKE3D model to
handle variable eccentricity worlds and to Tony Del Genio and the GISS ROCKE3D NExSS team for
comments and suggestions on this manuscript. Special thanks goes to the referee Dorian Abbott
for suggestions that improved the manuscript. The results reported herein benefited from
participation in NASA’s Nexus for Exoplanet System Science research coordination
network sponsored by NASA’s Science Mission Directorate.

This research has made use of NASA's Astrophysics Data System Bibliographic Services.


\begin{thebibliography}{}

\bibitem[Armstrong et al.(2004)]{Armstrong2004}
Armstrong, J C., Leovy, C.B. \& Quinn, T. 2004 Icarus, 171, 2, 255

\bibitem[Armstrong et al.(2014)]{Armstrong2014}
Armstrong J.C., Barnes R., Domagal-Goldman S., Breiner J., Quinn T.R. \&
Meadows V.S. 2014, Astrobiology, 14(4): 277-291.
doi:10.1089/ast.2013.1129.

\bibitem[Batygin \& Laughlin(2015)]{Batygin2015}
Batygin, K. \& Laughlin, G. 2015, PNAS, 112, 14, 4214,
doi:10.1073/pnas.1423252112

\bibitem[Bolmont et al.(2016)]{Bolmont2016}
Bolmont, E., Libert, A., Leconte, J. \& Selsis, F. 2016, \aa, 591, A106,
doi: 10.1051/0004-6361/201628073

\bibitem[Cowan et al.(2012)]{Cowan2012}
Cowan, N.B., Voigt, A. \& Abbot, D.S. 2012, \apj, 757, 80, doi:10.1088/0004-637X/757/1/80

\bibitem[Dressing et al.(2010)]{Dressing2010}
Dressing, C.D., Spiegel, D.S., Scharf, C.A., Menou, K. \& Raymond, S.N. 2010, \apj, 721, 1295,
doi:10.1088/0004-637X/721/2/1295

\bibitem[Ferreira et al.(2014)]{Ferreira2014}
Ferreira, D., Marshall, J. O'Gorman, P.A. \& Seager, S. 2014, \icarus, 236-248,
doi:10.1016/j.icarus.2014.09.015

\bibitem[Fritz et al.(2014)]{Fritz2014}
Fritz, J., Bitsch, B., Kuhrt, E., Morbidelli, A., Tornow, C., Wunnemann, K.,
Fernandes, V.A., Grenfell, J.L., Rauer, H., Wagner, R., Werner, S.C. 2014,
Planetary and Space Science, 98, 254, doi:10.1016/j.pss.2014.03.003


\bibitem[Gardner et al.(2006)]{Gardner2006}
Gardner, J. P., Mather, J. C., Clampin, M., et al. 2006, Space Sci Rev, 123, 485

\bibitem[Georgakarakos (2003)]{Georgakarakos2003}
Georgakarakos, N. 2003, \mnras, 345, 1, 340,	
doi:10.1046/j.1365-8711.2003.06942.x

\bibitem[Georgakarakos \& Eggl (2015)]{Georgakarakos2015}
Georgakarakos, N. \& Eggl, S. 2015, \apj, 802, 2, 94,
doi:10.1088/0004-637X/802/2/94

\bibitem[Georgakarakos et al.(2016)]{Georgakarakos2016}
Georgakarakos, N., Dobbs-Dixon, I., Way, M.J. 2016, \mnras, 461, 2, 1512,
doi:10.1093/mnras/stw1378

\bibitem[Godolt et al.(2015)]{Godolt2015}
Godolt, M., Grenfell, J. L., Hamann-Reinus, A., et al. 2015, Planet. Space Sci., 111, 62

\bibitem[Gilmozzi \& Spyromillio(2007)]{Gilmozzi2007}
Gilmozzi, R. \& Spryromillio, J. 2007, Messenger, 127, 11-19

\bibitem[Hu \& Yang(2014)]{HuYang2014}
Hu, Y. \& Yang, J. 2014, PNAS, 111, 2, 629, doi:10.1073/pnas.1315215111

\bibitem[Iro \& Deming(2010)]{Iro2010}
Iro, N. \& Deming, L.D. 2010, \apj, 712, 218, doi:10.1088/0004-637X/712/1/218

\bibitem[Kasting(1988)]{Kasting1988}
Kasting, J.F. 1988, Icarus, 74, 472–494.

\bibitem[Laskar et al.(1993)]{Laskar1993}
Laskar, J., Joutel, F., \& Robutel, P. 1993,
Nature 361, 6413, 615–-617, doi:10.1038/361615a0

\bibitem[Laskar et al.(2004)]{Laskar2004}
Laskar, J., Correia, A. C. M., Gastineau, M., Joutel, F., Levrard, B. \&
Robutel, P. 2004, Icarus 170, 343, doi:10.1016/j.icarus.2004.04.005

\bibitem[Legates \& Willmott(2008)]{Legates2008}
Legates, D.R. \& Willmott, C.J. 1990, Int. J. Climatol., 10: 111-127, doi:10.1002/joc.3370100202

\bibitem[Linsenmeier et al.(2015)]{Linsenmeier2015}
Linsenmeier, M., Pascale, M., Lucarini, V. 2015, Planetary and Space Science, 105, 43,
doi:10.1016/j.pss.2014.11.003

\bibitem[Lissauer et al.(2011)]{Lissauer2011}
Lissauer, J.J., Barnes, J.W. \& Chambers, J.E. 2011,
Icarus. 217: 77–87, doi:10.1016/j.icarus.2011.10.013

\bibitem[Milankovich(1941)]{Milankovich1941}
Milankovitch, M. 1941, Kanon der Erdbestrahlung und seine Anwendung auf das Eiszeitenproblem (Belgrade: Mihaila Curcica)

\bibitem[Peixoto \& Oort(1992)]{Peixoto1992}
Peixoto, J. \& Oort, A. 1992, Physics of Climate, Am Inst Phys, NY, 520 pp.

\bibitem[Pierrehumbert \& Ding(2016)]{PD2016}
Pierrehumbert, R.T. \& Ding, F. 2016, Proc of the Royal Soc of London A: Mathematical, Physical and Engineering Sciences, 472, 2190, 1364-5021, doi:10.1098/rspa.2016.0107

\bibitem[Ricker et al.(2014)]{Ricker2014}
Ricker, G. R., Winn, J. N., Vanderspek, R., et al. 2014, in P. SPIE, Vol. 9143, Space Telescopes and Instrumentation 2014: Optical, Infrared, and
Millimeter Wave, 914320

\bibitem[Rose(2015)]{Rose2015}
Rose, B.E.J. 2015, J. Geophys. Res. Atmos., 120, doi:10.1002/2014JD022659

\bibitem[Schmidt et al.(2014)]{Schmidt2014}
Schmidt, G. A., et al. 2014, J. Adv. Model. Earth Syst., 6, no.1, 141-184, doi:10.1002/2013MS000265.

\bibitem[Spiegel et al.(2008)]{Spiegel2008}
Spiegel, D. S., Menou, K. \& Scharf, C. A. 2008, ApJ, 681, 1609

\bibitem[Spiegel et al.(2010)]{Spiegel2010}
Spiegel, D.S., Raymond, S.N., Dressing, C.D., Scharf, C.A., \& Mitchell, J.L. 2010, \apj, 721:1308–1318.

\bibitem[Waltham(2004)]{Waltham2004}
Waltam, D. 2004, Astrobiology, 4, 4, 460

\bibitem[Ward \& Rudy(1991)]{Ward1991}
Ward, W.R. \& Rudy, D.J. 1991, Icarus, 94, 160

\bibitem[Way et al.(2017)]{Way2017}
Way, M.J. et al. 2016, to be submitted to \apjs

\bibitem[Williams \& Pollard(2002)]{WilliamsPollard2002}
Williams, D.M. \& Pollard, D. 2002, International Journal of Astrobiology 1:61,
doi:10.1017/S1473550402001064

\bibitem[Williams \& Pollard(2003)]{WilliamsPollard2003}
Williams, D.M. \& Pollard, D. 2003, International Journal of Astrobiology 2:1–19.
doi:10.1017/S1473550403001356

\bibitem[Yang et al.(2014)]{Yang2014}
Yang, J., Boue, G., Fabrycky, D.C. \& Abbot, D.S. 2013 \apjl, 787, L2,
doi:10.1088/2041-8205/787/1/L2

\end{thebibliography}
\end{document}